\documentclass[10pt,twocolumn,letterpaper]{article}

\usepackage{cvpr}              

\usepackage[pagebackref,breaklinks,colorlinks,citecolor=cvprblue]{hyperref}
\usepackage{multirow}
\usepackage{multicol}
\usepackage{adjustbox}
\usepackage[accsupp]{axessibility} 

\def\paperID{15573} 
\def\confName{CVPR}
\def\confYear{2024}

\title{PrPSeg: Universal Proposition Learning for Panoramic Renal Pathology Segmentation}

\author{Ruining Deng$^1$, Quan Liu$^1$, Can Cui$^1$, Tianyuan Yao$^1$, Jialin Yue$^1$, Juming Xiong$^1$, Lining Yu$^1$, Yifei\\
Wu$^1$, Mengmeng Yin$^2$, Yu Wang$^2$, Shilin Zhao$^2$, Yucheng Tang$^3$, Haichun Yang$^2$, Yuankai Huo$^{1,2}$*\\
\\
1. Vanderbilt University, 2. Vanderbilt Univeristy Medical Center, 3. NVIDIA Corp.\\
*contact author: yuankai.huo@vanderbilt.edu
}

\begin{document}

\definecolor{cvprblue}{rgb}{0.21,0.49,0.74}
\maketitle
\begin{abstract}
Understanding the anatomy of renal pathology is crucial for advancing disease diagnostics, treatment evaluation, and clinical research. The complex kidney system comprises various components across multiple levels, including regions (cortex, medulla), functional units (glomeruli, tubules), and cells (podocytes, mesangial cells in glomerulus). Prior studies have predominantly overlooked the intricate spatial interrelations among objects from clinical knowledge. In this research, we introduce a novel universal proposition learning approach, called panoramic renal pathology segmentation (PrPSeg), designed to segment comprehensively panoramic structures within kidney by integrating extensive knowledge of kidney anatomy.

In this paper, we propose (1) the design of a comprehensive universal proposition matrix for renal pathology, facilitating the incorporation of classification and spatial relationships into the segmentation process; (2) a token-based dynamic head single network architecture, with the improvement of the partial label image segmentation and capability for future data enlargement; and (3) an anatomy loss function, quantifying the inter-object relationships across the kidney.

\end{abstract}    
\vspace{-5mm}
\section{Introduction}
\label{sec:intro}

\begin{figure*}[t]
\centering 
\includegraphics[width=1.0\linewidth]{{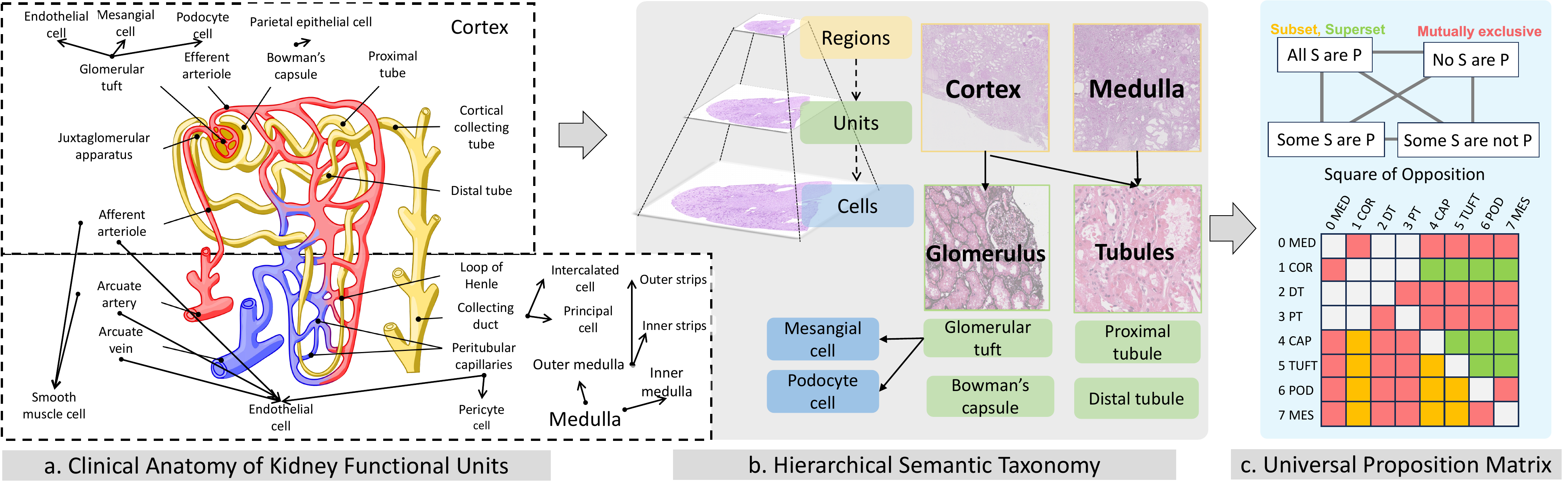}}
\caption{
\textbf{Knowledge transformation from kidney anatomy to computational modeling -- }
This figure demonstrates the transformation of intricate clinical anatomical relationships within the kidney into a structured computational matrix. (a) Pathologists examine histopathology following the kidney anatomy. (b) This study revisits such kidney anatomy with hierarchical semantic taxonomy. (c) The proposed PrPSeg method further mathematically abstracts the semantic taxonomy as a universal proposition matrix. This matrix serves as a foundation for our computational model, reflecting the complex interplay of anatomical elements in the kidney.} 
\label{Fig:tease} 
\end{figure*}

 Digital pathology has revolutionized the field of pathology~\cite{bandari2016renal}, not only facilitating the transition from local microscopes to remote monitoring for pathologists but also providing a significant opportunity for large-scale computer-assisted quantification in pathology~\cite{bengtsson2017computer, marti2021digital, gomes2021building}. In the clinical anatomy of renal pathology, there are different levels of quantification necessary for disease diagnosis~\cite{mounier2002cortical}, severity recognition~\cite{kellum2008acute}, and treatment effectiveness evaluation~\cite{jimenez2006mast}, ranging from region-level objects (like medulla and cortex) to functional units (glomerulus, tubules, vessels, etc.) and even to individual cells within these units. These tasks, especially at the functional unit and cell levels, are prone to errors and variability in human examination and require labor-intensive efforts~\cite{wijkstrom2018morphological, zheng2021deep, imig2022interactions}. Therefore, achieving comprehensive quantification from regions to cells is necessary in renal pathology, but it remains an inevitably laborious task with manual human effort.

While many studies have developed pathological image segmentation techniques for pixel-level tissue characterization, particularly using deep learning methods~\cite{kumar2017dataset, ding2020multi, ren2017computer, bel2018structure, zeng2020identification}, they still encounter three major limitations: (1) Current multi-network and multi-head designs~\cite{jayapandian2021development, li2019u, hermsen2019deep, zhang2021dodnet, wu2022tgnet, deng2023omni} focus only on single tissue structures or structures at similar scales, lacking a comprehensive approach to achieve all-encompassing segmentation across different levels from regions to cells. (2) These approaches require modifications to their architectures when new classes are introduced, preventing the reuse of existing backbones without significant alterations; (3) Comprehensive semantic (multi-label) segmentation and quantification on renal histopathological images remain challenging due to the intricate spatial relationships among different tissue structures. The spatial relationships between these different objects are illustrated in~\Cref{Fig:tease}.  \emph{Understanding these interrelations in renal pathology is crucial for achieving effective all-encompassing segmentation, yet recent advancements in deep learning have not fully incorporated this comprehensive modeling into the training process, nor have they achieved panoramic segmentation for the complete anatomy of the kidney.}

To address these challenges, we introduce a token-based dynamic head network designed to achieve panoramic renal pathology segmentation (PrPSeg) by modeling the spatial relationships among all objects. This approach allows for the reuse of the same architectural framework, even when the dataset size expands. A universal proposition matrix is established to translate anatomical relationships into computational modeling concepts. A anatomy loss is also proposed to integrate spatial relationships into the model training as a form of semi-supervised learning. To our knowledge, this is the first deep learning algorithm to accomplish panoramic segmentation in renal pathology, demonstrating superior performance in all-encompassing segmentation.

The contributions of this paper are threefold: 
\begin{itemize}
    \item The design of a comprehensive universal proposition matrix for renal pathology, facilitating the incorporation of classification and spatial relationships into the segmentation process.
    \item The development of a token-based dynamic head in a single network architecture, improving partial label image segmentation.
    \item The formulation of an anatomy loss function, quantifying the inter-object relationships across the kidney.

\end{itemize}
\section{Relative Work}
\label{sec:relativework}

\subsection{Renal Pathology Segmentation}
\label{subsec:segmentation}
Recent advancements in deep learning have positioned Convolutional Neural Networks (CNNs) and Transformer-based networks as leading methods for image segmentation, particularly in renal pathology~\cite{feng2022artificial,hara2022evaluating}. Innovations in this field have ranged from CNN cascades for sparse tissue segmentation~\cite{gadermayr1708cnn} to the deployment of AlexNet for pixel-wise classification and detection~\cite{gallego2018glomerulus}. Notably, multi-class learning approaches using SegNet-VGG16 and DeepLab v2 have been implemented for detecting various glomerular structures and renal pathologies~\cite{bueno2020glomerulosclerosis,lutnick2019integrated}. In addition, instance segmentation and Vision Transformers (ViTs) have begun to find applications in this domain~\cite{johnson2019automatic,nguyen2021evaluating, gao2021instance,yan2023self}.

However, most existing methods focus on segmenting single tissue types or multiple structures at similar levels, such as glomeruli and tubules~\cite{gupta2018iterative,kannan2019segmentation, marechal2022automatic}. Comprehensive approaches capable of spanning from tissue region level to cell level remain unexplored. Moreover, some methods prioritize disease-positive region segmentation over a holistic understanding of kidney morphology~\cite{jing2022segmentation,lin2022adversarial}.

Recent methods utilize hierarchical information for semantic segmentation~\cite{Li_2022_CVPR,ke2022hsg} or classification and prediction~\cite{chen2022scaling}. However, these methods primarily focus on class-based relationships between objects. While all objects have a uniform resolution in natural images, this approach neglects the emphasis on pixel-wise anatomical and spatial relationships at multiple resolutions in kidney datasets.

Building upon these insights, our work introduces a token-based approach, leveraging class-specific and scale-specific tokens. This method is designed to capture heterogeneous features and employs semi-supervised learning to understand pixel-wise spatial relationships across multiple scales, achieving panoramic segmentation in renal pathology.

\subsection{Dynamic Single Network}
\label{subsec:dynamicnetwork}

\begin{figure*}[t]
\centering 
\includegraphics[width=1\linewidth]{{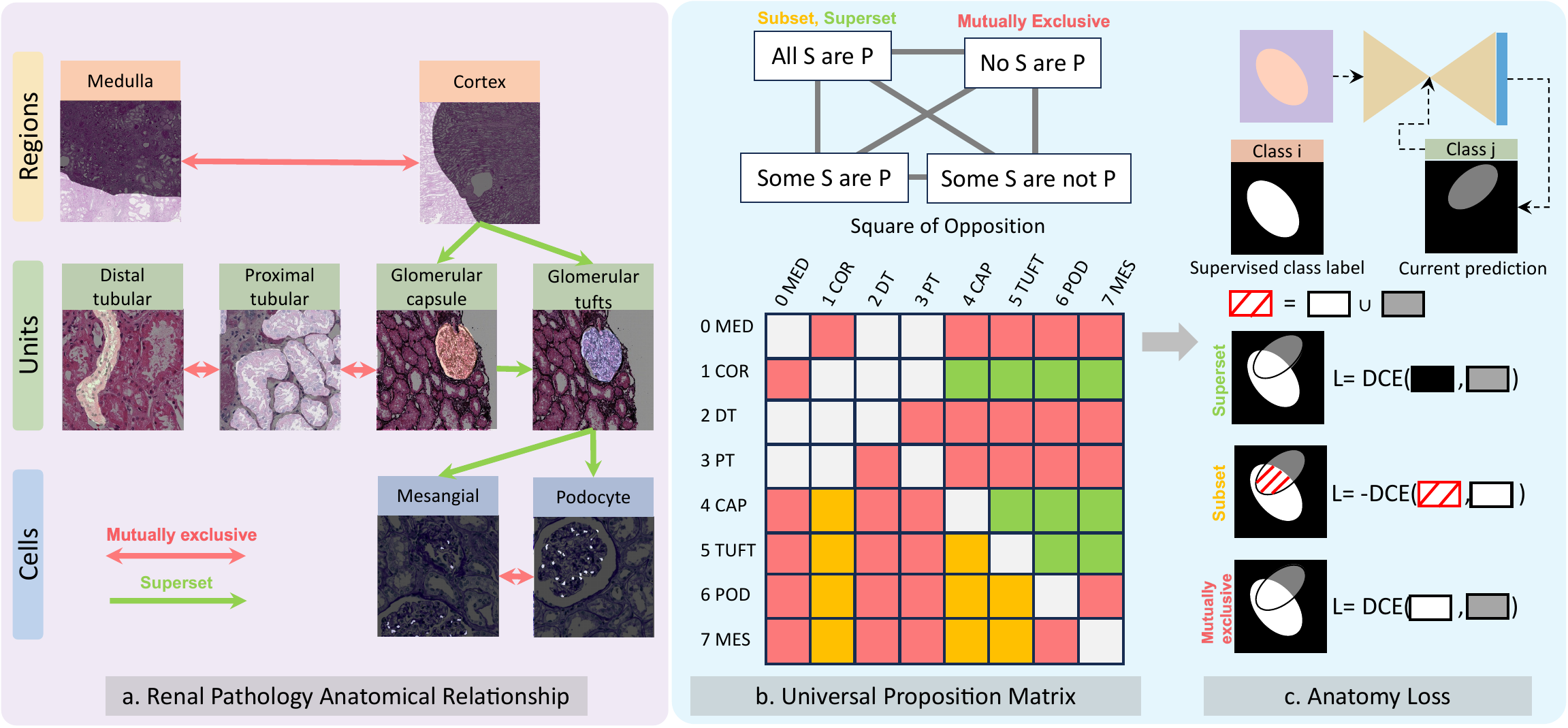}}
\caption{
\textbf{Universal proposition matrix with anatomy loss -- }
This figure shows the key innovation of the proposed method. (a) Multi-scale (region-level, unit-level, and cell-level) hierarchical semantic taxonomy is presented. (b) The proposed PrPSeg mathematically models the semantic taxonomy as a universal proposition matrix, which delineates robust constraints and relationships between anatomical entities. (c) We further encode the universal proposition matrix as a novel anatomy loss function, designed to operationalize the affirmative and negatory relationships inherent in kidney anatomy.}
\label{fig.anatomy} 
\end{figure*}

While multi-head single network designs have been proposed for multi-class renal pathology segmentation~\cite{gonzalez2018multi,fang2020multi,bouteldja2021deep,chen2019med3d}, they often require dense multi-class annotations. Given the labor-intensive nature of such annotations, pathology data is frequently partially labeled. Additionally, forming spatial correlations, such as subset/superset relationships, remains a challenge in multi-class uplsegmentation frameworks.

Recent developments in dynamic neural networks have paved the way for more comprehensive segmentation using single multi-label networks, even with partially labeled data~\cite{zhang2021dodnet,deng2023omni}. These networks dynamically generate neural network parameters, adapting to various imaging contexts. However, they predominantly rely on binary segmentation approaches and do not fully integrate spatial correlations in their training processes.

Our work extends these concepts by translating spatial correlations from anatomy into a programming model, represented as a matrix coupled with an anatomy-based loss for semi-supervised learning. This strategy effectively harnesses the affirmative and negatory relationships in kidney anatomy, enabling detailed and comprehensive segmentation. It enhances the model’s ability to distinguish and classify the complex structures within kidney, representing a significant advancement in the field.

\section{Method}
\label{sec:method}

\subsection{Problem Formulation}
\label{subsec:problem}
This study aims to segment an array of anatomical concepts in renal pathology, encapsulating three conceptual layers — regions ($R_1, R_2$), functional units ($F_1, F_2, F_3, F_4$), and cells ($C_1, C_2$) — spanning 8 distinct objects.

Leveraging anatomical learning, our approach is tailored to achieve comprehensive segmentation in renal pathology by effectively interpreting both affirmative and negatory relationships within the anatomical relationship.

\begin{figure*}[t]
\centering 
\includegraphics[width=1\linewidth]{{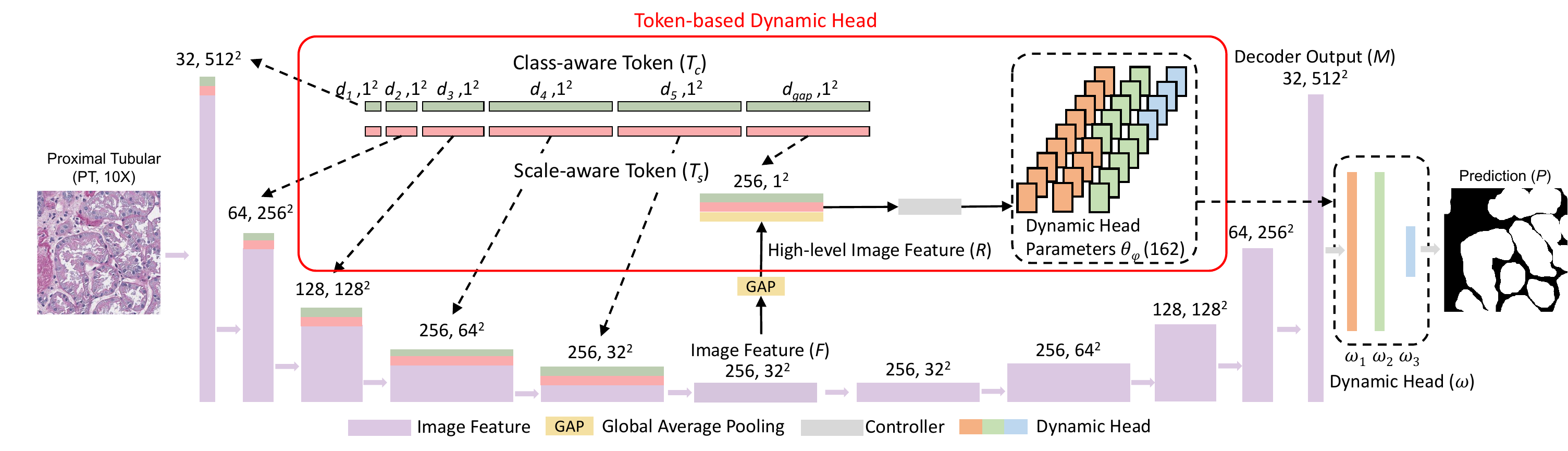}}
\caption{
\textbf{Token-based dynamic head network architecture -- }
This figure illustrates the architecture of the proposed PrPSeg method. It incorporates a residual U-Net backbone, augmented with class-aware and scale-aware tokens. These tokens are integrated into each block of the encoder, as well as the Global Average Pooling (GAP) block, ensuring a comprehensive understanding of both class and scale features. Such features are aggregated by a fusion block to adaptively generate the parameters for a single dynamic segmentation head. The proposed method is able to segment all hierarchical semantic anatomies using a single network.} 
\label{Fig.architecture} 
\end{figure*}

The pipeline is composed of three integral components:

\noindent\textbf{Universal proposition matrix:} A meticulously crafted universal proposition matrix is employed to elucidate the anatomical relationships among various objects. This matrix aids in enhancing structural understanding from an engineering perspective, facilitating better cognition of the complex renal architecture.

\noindent\textbf{Token-based dynamic head backbone:} We have developed a token-based dynamic head backbone architecture that is adept at interpreting class-aware and scale-aware knowledge pertinent to renal pathology. This component is pivotal in accommodating the extensibility requirements posed by the introduction of new data, ensuring the model's adaptability and scalability.

\noindent\textbf{Anatomy loss function:} A novel anatomy loss function has been formulated, which operationalizes the affirmative and negatory relationships inherent in kidney anatomy. This function is a critical element in achieving nuanced, all-encompassing segmentation, bolstering the model's ability to discern and categorize the intricate structures within the kidney.

\subsection{Universal proposition matrix}
\label{subsec:matrix}
Renal pathology encompasses regions (the medulla and cortex), functional units (glomerulus, tubules, etc.), with corresponding cellular structures. Pathologists analyze the morphology of the kidney by examining the functional flow through these heterostructures. Traditionally, each heterostructure undergoes isolated examination and quantification to meet specific demands, often overlooking the homostructures within each unit, such as podocyte cells, mesangial cells in glomerular tufts. To enhance the understanding of the integrated kidney structure, we propose transforming kidney anatomy into an anatomical relationship map, using principles of affirmation and negation from linguistic and grammatical concepts.

To translate these anatomy concepts into an engineering framework, we adopt Aristotle's logic theory to develop an anatomical map for renal pathology. Aristotle's theory examines the relationships between objects using four fundamental categorical propositions. Upon closer inspection, complex propositions reveal themselves as collections of simpler claims derived from these initial propositions. Specifically, we utilize two terms from Aristotle's theory: \textit{(1) Universal Affirmation: ``All S are P," and (2) Universal Negation: ``No S are P,"}, to universally assert properties for all group members, indicating strong constraints and relationships. For example, in the context of two kidney structures, $A$ and $B$, if $B$ is within $A$, $B$ is a subset of $A$, following the rule of Universal Affirmation. Conversely, if $A$ and $B$ have no inclusion relationship, they are mutually exclusive, aligning with Universal Negation. These propositions are employed to construct an anatomical relationship map representing the classification and spatial relationships among renal pathology objects, as illustrated in~\Cref{fig.anatomy}.

This anatomical relationship is characterized by:

\noindent\textbf{Uniqueness:} Each pair of objects is linked by a single proposition. The expanding structure of the map, devoid of cycles, ensures stable inheritance relationships from regions down to cellular levels.

\noindent\textbf{Transmissibility:} Indirect relationships between objects can be deduced from direct relationships, as established by the two fundamental categorical propositions. Relationships between objects not directly connected can be determined by combining propositions along their connecting path.

Following the translation of this knowledge from clinical anatomy to an engineering paradigm, we introduce a universal proposition matrix, $M_t\in\mathbb{R}^{n \times n}$, to facilitate implementation in computational models. Here, $n$ represents the number of classes within the map. The matrix values are defined in~\Cref{eq:matrix}:

\vspace{-5mm} 
\begin{equation}
    M_t(i,j) = 
    \begin{cases}
    1, \quad if I  \subseteq J \\
    -1, \quad if I \supseteq J \\
    2, \quad if I \cap J = \emptyset \\
    0, \quad \text{otherwise}
    \end{cases} 
    i,j = 1,2,...,n
\label{eq:matrix}
\end{equation}

This matrix is designed to be extendable with the introduction of new data.

\subsection{Token-based Dynamic Head Network}
\label{subsec:token}

\begin{figure}
\centering 
\includegraphics[width=1\linewidth]{{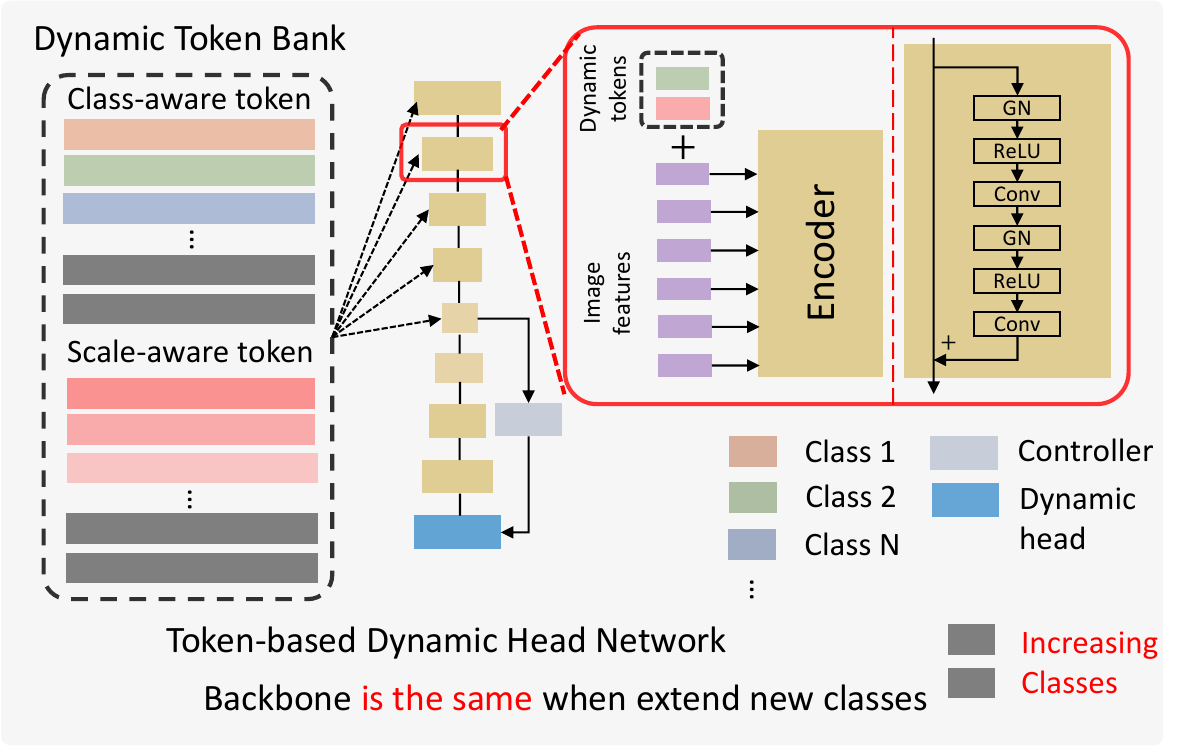}}
\caption{
\textbf{Token-based dynamic head -- }
This figure visualizes the architecture of our proposed token-based dynamic head backbone. Central to our design is the ability to maintain a consistent model architecture while dynamically accommodating an increasing number of segmentation classes. This flexibility is achieved by extending the dimensions of the tokens, rather than altering the backbone structure. Key components include a dynamic token bank with class-aware and scale-aware tokens, an encoder, and a dynamic head network, all orchestrated to efficiently handle class expansion without necessitating changes to the backbone.} 
\label{Fig.token} 
\end{figure}

Pathological image segmentation faces three main challenges: (1) Heterogeneous object annotations are often partially labeled, with only one type of tissue annotated per pathological image; (2) It is challenging to form anatomical relationships (e.g. subset/superset relationships) in multi-class segmentation; for example, it is difficult to simultaneously segment glomerular capsule, tufts, and cells that share regions; (3) The annotation process on giga-pixel images is labor-intensive, leading to an ongoing data collection process with class extension. Therefore, a segmentation backbone optimized for binary segmentation of multiple classes with spatial overlap, and adaptable to data enlargement, is required. Previous designs, including multi-head and dynamic-head, are suboptimal for partially labeled learning and data extension, leading to changes in the backbone architecture and insufficient performance.

In this work, we propose a token-based dynamic head backbone designed to maintain consistent model architecture while accommodating a possibly increasing number of segmentation classes (In~\Cref{Fig.token}). The backbone can be supervised with partially labeled data, understanding anatomical relationships effectively. The entire architecture is demonstrated in~\Cref{Fig.architecture}. The backbone of our proposed method is the Residual-U-Net from Omni-Seg~\cite{deng2023omni}, chosen for its superior segmentation performance in renal pathology. Instead of using dimensionally-changeable one-hot vectors for class-aware encoding and scale-aware encoding, we use dimensionally-stable class-aware tokens ($T_c\in\mathbb{R}^{n \times d}$) and scale-aware tokens ($T_s\in\mathbb{R}^{4 \times d}$) to comprehend the contextual information in the encoder ($E$) of the model. Here, $d$ is the sum of the channel numbers of blocks ($d_1+ d_2+\dots+d_b+d_{gap}$) in the encoder. Each interval of the channels represents level-specific features in each level of the encoder. Each class has a one-dimensional token $t_c\in\mathbb{R}^{1 \times d}$ to store class-specific knowledge at the feature level among the whole dataset, while each magnification has a one-dimensional token $t_s\in\mathbb{R}^{1 \times d}$ for providing scale-specific knowledge across four scales (5$\times$, 10$\times$, 20$\times$, and 40$\times$). Inspired by the Vision Transformer~\cite{dosovitskiy2020image}, for an image $I$ of class $i$ with magnification $m$, at the $b$-th encoder block, the corresponding class token $T_c(i)$ and $T_s(m)$ are added to the current feature map ($e_{b-1}$) before being fed into the current block ($E_b$). This process is defined by the following equation:

\begin{equation}
    \text{Ind}_{start} = \sum_{k=1}^{b-1} d_k,  \text{Ind}_{end} = \sum_{k=1}^{b} d_k
\label{eq:fusion}
\end{equation}

\begin{equation}
\begin{aligned}
    e_b = E_b(&  T_c[i][\text{Ind}_{start}:\text{Ind}_{end}] + \\
    &   T_s[m][\text{Ind}_{start}:\text{Ind}_{end}] + e_{b-1})
\end{aligned}
\label{eq:token}
\end{equation}

With the class token $T_c$ and scale token $T_s$, the encoder captures domain-specific features in the image.

To further integrate class-specific and scale-specific information into the embedded features, we combine the last intervals of the class-token and scale token into the low-dimensional feature embedding at the bottom of the Residual-U-Net architecture. The image feature $F$ is summarized by Global Average Pooling (GAP) and transformed into a feature vector in $\mathbb{R}^{d_{gap}}$. $d_{gap}$ is the dimension of GAP feature. Differing from Omni-Seg~\cite{deng2023omni}, which implements a triple outer product to combine three vectors into a one-dimensional vector via a flatten function, we use a single 2D convolutional layer controller, $\varphi$, as a feature fusion block to refine the fusion vector as the final controller for dynamic head mapping:

\vspace{-3mm}
\begin{equation}
    \omega = \varphi(\text{GAP}(F)+T_c[i][:-d_{gap}]+T_s[i][:-d_{gap}];\Theta_\varphi)
\label{eq:fusion}
\end{equation}

\noindent where $\text{GAP}(F)$, $T_c$, and $T_s$ are combined by the addition operation, and $\Theta_\varphi$ represents the number of parameters in the dynamic head.

Following the approach of~\cite{deng2023omni}, a binary segmentation network is used to achieve multi-label segmentation via a dynamic filter. From the multi-label multi-scale modeling described above, we derive joint low-dimensional image feature vectors, class-specific tokens, and scale-specific tokens at an optimal segmentation magnification. These are then mapped to control a lightweight dynamic head, specifying (1) the target tissue type and (2) the corresponding pyramid scale.

The dynamic head consists of three layers. The first two layers contain eight channels each, while the final layer comprises two channels. We directly map parameters from the fusion-based feature controller to the kernels in the 162-parameter dynamic head to achieve precise segmentation from multi-modal features. The filtering process is expressed by the following equation:

\vspace{-2mm}
\begin{equation}
    P = ((((M \cdot \omega_1) \cdot \omega_2) \cdot \omega_3)
\label{eq:dynamic-head}
\end{equation}

\noindent where $\cdot$ denotes convolution, $P\in\mathbb{R}^{2\times W \times H}$ is the final prediction, and $W$ and $H$ correspond to the width and height of the dataset, respectively.

The benefits of the dynamic-token design are twofold: (1) The backbone architecture remains stable and reusable as new classes are introduced, which is advantageous for incremental learning. (2) The binary segmentation scheme allows the model to predict multiple classes with spatial overlap, outperforming other multi-head and dynamic-head designs.

\subsection{Anatomy Loss Function}
\label{subsec:loss}

With the introduction of the universal proposition matrix and token-based dynamic head architecture, we propose an online semi-supervised anatomy learning strategy to incorporate spatial correlation into the training process for comprehensive segmentation. For a given image $I$ with a labeled class $i$, represented as $Y_i$, we generate predictions $Y'_j$ for another class $j$ on the same image. We then use the anatomical relationship to supervise the correlation between the supervised label $Y_i$ and the semi-supervised prediction $Y'_j$: (1) If $i$ is a superset of $j$, $Y'_j$ should not exceed the region of $Y_i$; conversely, (2) if $i$ is a subset of $j$, $Y'_j$ should cover $Y_i$ as comprehensively as possible; and (3) if $i$ is mutually exclusive with $j$, the overlap between $Y_i$ and $Y'_j$ should be minimized. The total anatomy loss is defined by the following equations:



\vspace{-5mm}
\begin{equation}
    L_{\text{upl}}(i,j) = 
    \begin{cases}
    \text{DCE}(1 - Y_i, Y'_j), \quad \text{if} \quad M_t(i,j) = 1 \\
    - \text{DCE}(Y_i, Y_i \cup Y'_j ), \quad \text{if} \quad M_t(i,j) = -1 \\
    \text{DCE}(Y_i, Y'_j), \quad \text{if} \quad M_t(i,j) = 2 \\
    0, \quad \text{if} \quad M_t(i,j) = 0 \\
    \end{cases} 
\label{eq:anatomy_loss}
\end{equation}
\vspace{-3mm}

\noindent where $M_t$ is the anatomy matrix and $\text{DCE}$ denotes the Dice Loss. The total loss function is an aggregate of supervised and semi-supervised losses, weighted by $\lambda_{\text{upl}}$.

\vspace{-2mm}
\begin{equation}
\begin{aligned}
    L(i) = & \text{DCE}(Y_i,Y'_i) + \text{BCE}(Y_i,Y'_i)\\
    & + \lambda_{\text{upl}} \sum_{j=1}^{n}L_{\text{upl}}(i,j) \quad ( j \neq i )
\end{aligned}
\label{eq:loss_function}
\end{equation}
\vspace{-5mm}

\noindent where $\text{BCE}$ represents the Binary Cross-Entropy loss, and $Y'_i$ is the prediction for class $i$.

\section{Data and Experiment}
\label{sec:details}

\subsection{Data}
\label{subsec:data}

Our model leverages an 8-class, partially labelled dataset spanning various biological scales, from regions to cells. The dataset's structure is detailed in \Cref{tab:dataset}. We sourced the human kidney dataset from three distinct resources:

\noindent\textbf{Regions:} Whole slide images of wedge kidney sections stained with periodic acid-Schiff (PAS, n=138) were obtained from from non-cancerous regions of nephrectomy samples. The samples were categorized into several groups based on clinical data, including normal adults (n=27), patients  with hypertension (HTN, n=31), patients with diabetes (DM, n=4), patients with both hypertension and diabetes (n=14), normal aging individuals (age$>$65y, n=10), individuals with aging and hypertension (n=36), and individuals with aging, hypertension, and diabetes (n=16). These tissues were scanned at 20$\times$ magnification and manually annotated in QuPath~\cite{bankhead2017qupath}, delineating medulla and cortex contours. The WSIs were downsampled to 5$\times$ magnification and segmented into 1024$\times$1024 pixel patches. Corresponding binary masks were derived from the contours.

\noindent\textbf{Functional Units:} Using 459 WSIs from NEPTUNE study~\cite{barisoni2013digital}, encompassing 125 patients with minimal change disease, we extracted 1,751 Regions of Interest (ROIs). These ROIs were manually segmented to identify four kinds of morphology objects with normal structure and methodology outlined in~\cite{jayapandian2021development}. Each image, at a resolution of 3000$\times$3000 pixels (40$\times$ magnification, 0.25 $\mu m$ per pixel), represented one of four tissue types stained with Hematoxylin and Eosin Stain(H\&E), PAS, Silver Stain (SIL), and Trichrome Stain (TRI). We treated these four staining methods as color augmentations and resized the images to 256$\times$256 pixels, maintaining the original data splits from~\cite{jayapandian2021development}.

\noindent\textbf{Cells:} We acquired 11 PAS-stained WSIs from nephrectomy specimens with normal kidney function and morphology, scanned at 20$\times$ magnification. These pathological images were cropped into 512$\times$512 pixel segments to facilitate cell labeling, following the annotation process described in~\cite{deng2023democratizing}.

The dataset was partitioned into training, validation, and testing sets at a 6:1:3 ratio across all classes, with splits conducted at the patient level to prevent data leakage.

\subsection{Experiment Details}
\label{subsec:experiment}

\begin{table}
\begin{center}
\caption{Data collection}


\begin{tabular}{l|cccc}
\toprule
Class  & Patch \#  & Size & Scale & Stain\\
\midrule
Medulla & 1619 &  1024$^{2}$ & 5 $\times$ & P \\
Cortex & 3055 & 1024$^{2}$ & 5 $\times$ & P\\
\midrule
DT & 4615 & 256$^{2}$ & 10 $\times$ & H,P,S,T \\
PT & 4588 &  256$^{2}$ & 10 $\times$ & H,P,S,T \\
Cap. & 4559  &  256$^{2}$ & 5 $\times$ & H,P,S,T  \\
Tuft &  4536 &  256$^{2}$ & 5 $\times$ & H,P,S,T \\
\midrule
Pod. & 1147  & 512$^{2}$ & 20 $\times$ & P \\
Mes. & 789  & 512$^{2}$ & 20 $\times$ & P \\
\bottomrule
\end{tabular}
\label{tab:dataset} 
\end{center}
\text{*DT is distal tubular; PT is proximal tubular; }\\
\text{*Cap. is glomerular capsule; Tuft is glomerular tuft;}\\
\text{*Pod. is podocyte cell; Mes. is mesangial cell.}\\
\text{*H is H$\&$E; P is PAS; S is SIL; T is TRI.}
\vspace{-4mm}
\end{table}

\begin{table*}
\begin{center}
\caption{
Performance of deep learning based multi-class panoramic segmentation. Dice similarity coefficient scores (\%) are reported.}
\begin{adjustbox}{width=1.0\textwidth}
\begin{tabular}{l|l|ccccccccc}
\toprule
\multirow{2}{0.8in}{Method} & \multirow{2}{0.8in}{Backbone} & \multicolumn{2}{c}{Regions} & \multicolumn{4}{c}{Functional units} & \multicolumn{2}{c}{Cells} & \multirow{2}{0.8in}{Average} \\
\cmidrule(lr){3-4}
\cmidrule(lr){5-8}
\cmidrule(lr){9-10}
 & & Medulla & Cortex & DT & PT  & Cap. & Tufts & Pod. & Mes. & \\
\midrule
U-Nets~\cite{gonzalez2018multi} & CNN & 23.86 & 66.42 & 47.61 & 51.04 & 45.36 & 46.62 & 49.92 & 49.87 & 47.58 \\
DeepLabV3~\cite{lutnick2019integrated}& CNN & 41.70 & 61.26 & 63.92 & 65.31 & 72.82 & 81.93 & 49.92 & 49.87 & 60.84 \\
Residual-U-Net~\cite{salvi2021automated} & CNN& 13.21 & 69.97 & 67.03 & 76.59 & 70.58 & 82.37 & 63.99 & 64.54 & 63.54 \\
Multi-kidney~\cite{bouteldja2021deep}& CNN & 13.13 & 69.77 & 61.58 & 62.13 & 82.27 & 62.03 & \textbf{71.82} & 65.13 & 60.98 \\
Omni-Seg~\cite{deng2023omni} & CNN & \textbf{72.96} & 71.84 & 69.76 & 81.48 & 92.31 & 92.36 & 67.28 & 65.31 & 76.66 \\
\midrule
Segmenter~\cite{strudel2021segmenter}& Transformer & 56.38 & 67.34 & 54.81 & 69.14 & 67.16 & 66.78 & 49.92 & 49.87 & 60.18 \\
SegFormer~\cite{xie2021segformer} & Transformer & 54.82 &  67.68 &  62.65 & 75.87 &  77.46 &  60.42 & 62.43 & 60.21 & 65.19 \\
Unetr~\cite{hatamizadeh2022unetr} & Transformer & 16.22 & 69.70 &  61.99 & 69.35 & 72.48 & 58.10 & 58.14 & 56.32 & 57.79 \\
Swin-Unetr~\cite{hatamizadeh2021swin} & Transformer & 13.81 & 68.92 & 70.76 & 76.93 & 81.28 & 72.92 & 49.92 & 64.46 & 62.38 \\
\midrule
PrPSeg (Ours) & CNN &  72.38 &  \textbf{72.64} & \textbf{72.45} &  \textbf{85.27} &  \textbf{94.23} &  \textbf{94.40} & 70.98 & \textbf{66.96} & \textbf{78.66} \\
\bottomrule

\end{tabular}
\end{adjustbox}
\end{center}
\text{*DT is distal tubular; PT is proximal tubular; Cap. is glomerular capsule}\\
\text{*Pod. is glomerular podocyte cell; Mes. is glomerular mesangial cell}\\
\vspace{-4mm}
\label{tab:Qualitative}
\end{table*}

\begin{figure*}[t]
\centering 
\includegraphics[width=1\linewidth]{{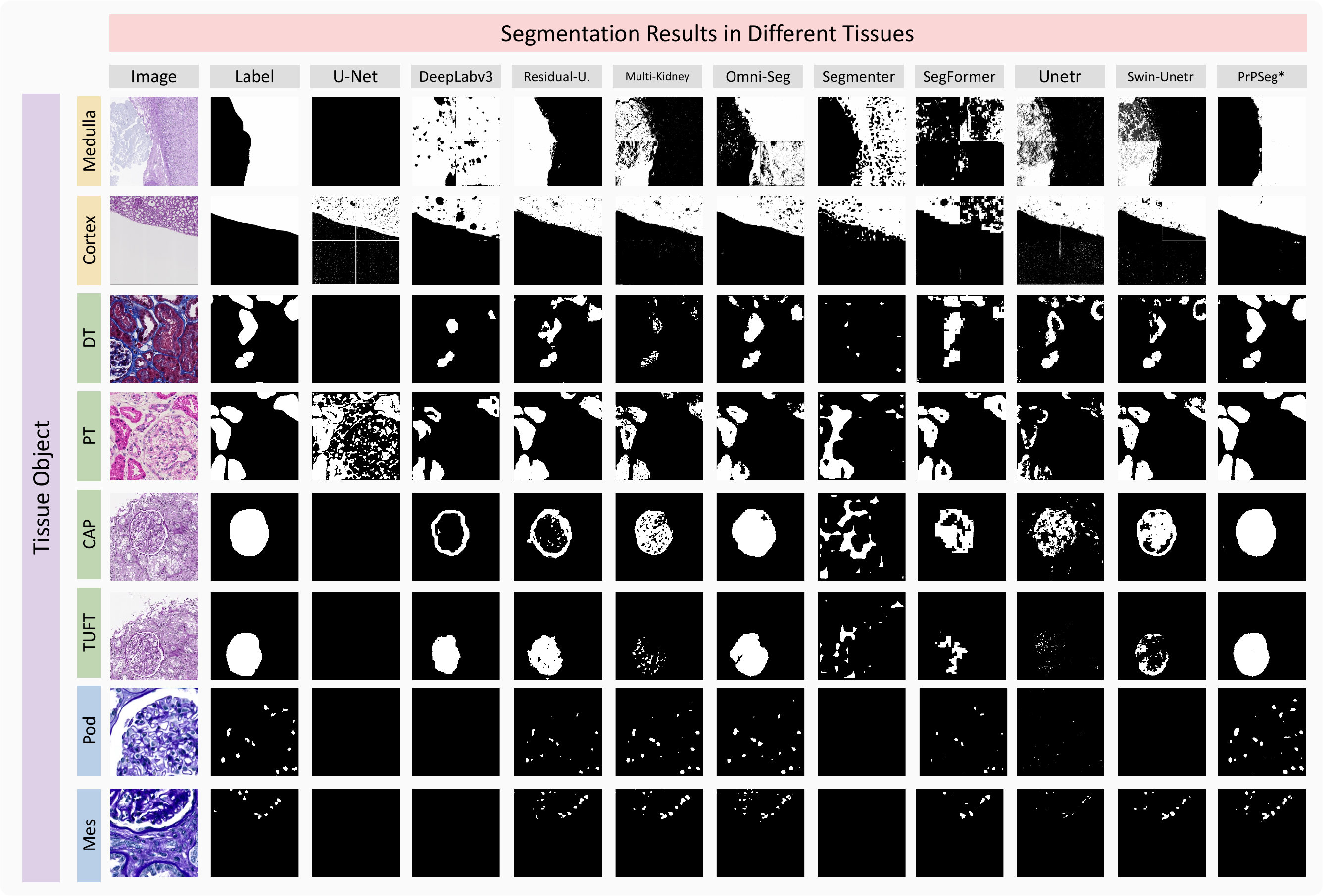}}
\caption{
\textbf{Validation qualitative results -- } This figure shows the qualitative results of different approaches. The proposed method achieved superior panoramic renal pathology segmentation on 8 classes range regions to cells with fewer false positives, false negatives, and morphological errors.} 

\label{Fig.Qualitative} 
\end{figure*}

The training process of our model was divided into two distinct phases. In the initial phase, spanning the first 50 epochs, we employed a supervised learning strategy focused on minimizing binary dice loss and cross-entropy loss. Subsequently, for the remaining epochs, both supervised and semi-supervised learning strategies were utilized, incorporating anatomy loss to explore the spatial correlation among multiple objects. In our experiments, $\lambda_{\text{upl}}$ is 0.1.

All images were either randomly cropped or padded to a uniform size of \( 512 \times 512 \) pixels prior to being fed into the model in the training stage. We established 8 separate image pools, each designated for different tissue types, to organize training batches. This approach follows the image pooling strategy from Cycle-GAN~\cite{zhu2017unpaired}. The batch size was set to 4, while each image pool could accommodate up to 8 images. Once an image pool accumulated more than the batch size, the images were retrieved from the pool and input into the network for processing. During each backpropagation step, Binary Dice Loss and Cross-entropy Loss were combined as the loss function in the supervised learning phase.

For weight optimization, we employed the Adam optimizer with an initial learning rate of 0.001 and a decay factor of 0.99. We also implemented general data augmentation techniques, including Affine transformations, Flip, Contrast adjustment, Brightness adjustment, Coarse Dropout, Gaussian Blur, and Gaussian Noise. These augmentations, sourced from the imgaug package~\cite{imgaug}, were applied to the entire training dataset with a probability of 0.5.

Model selection was based on the mean Dice coefficient score across 8 classes, evaluated on the validation dataset. The best-performing models within the first 100 epochs were then assessed on the testing dataset. Testing images were initially processed using either center-cropping or non-overlapping tiling to attain the same uniform size \( 512 \times 512 \) pixels. For evaluation, these images were subsequently either center-cropped or re-aggregated to their original dimensions. All experiments were conducted on a uniform platform, specifically a workstation equipped with an NVIDIA RTX A6000 GPU.

\section{Result}
\label{sec:results}

We conducted a comparative analysis of our proposed Universal Proposition Learning approach against various baseline models. These models include multi-class segmentation architectures such as (1) U-Nets~\cite{gonzalez2018multi}, (2) DeepLabV3~\cite{lutnick2019integrated}, (3) Residual-U-Net~\cite{salvi2021automated}, (4) a CNN-based multi-class kidney pathology model~\cite{bouteldja2021deep}, (5) Omni-Seg~\cite{deng2023omni}, (6) Segmenter-ViT/16~\cite{strudel2021segmenter}, (7) SegFormer~\cite{xie2021segformer}, (8) Unetr~\cite{hatamizadeh2022unetr}, and (9) Swin-Unetr~\cite{hatamizadeh2021swin}.

\subsection{Panoramic Segmentation Performance}
\label{subsec:segmentationresult}

Table \textcolor{red}{2} and \Cref{Fig.Qualitative} showcase the results from an 8-class segmentation evaluation. Table \textcolor{red}{2} demonstrates that our proposed method, PrPSeg, surpasses baseline models in most evaluated metrics. \Cref{Fig.Qualitative} further highlights the qualitative superiority of our approach, evidenced by reduced instances of false positives, false negatives, and morphological errors. The Dice similarity coefficient (Dice: \%, the higher, the better) was employed as the primary metric for quantitative performance assessment. The results indicate that, while multi-head designs struggle with managing spatial relationships between objects (e.g., subset/superset relationships between the capsule and tuft), the dynamic-head paradigm exhibits superior performance compared to other methods.

\subsection{Ablation Study}
\label{subsec:ablationstudy}
\Cref{tab:differentdesigns} showcases the enhancements brought about by our proposed token-based design and learning strategy across two different backbone architectures. The results indicate that the token-based dynamic head design boosts the model's performance in segmenting all levels of objects. With the integration of the token-based dynamic head architecture and universal proposition learning, the proposed method exhibited superior performance across all considered metrics.

We also provide an ablation study for two data extension scenarios in the Appendix (\textcolor{red}{A.1}). The proposed PrPSeg method is flexible to extend to new classes by merely updating tokens and the adaptable proposition matrix, without changing the backbone network, while demonstrating superior performance compared to baseline methods across all seven new classes.

\begin{table}
\caption{Ablation study of different design. Dice similarity coefficient scores (\%) are reported.}
\begin{center}
\begin{adjustbox}{width=0.45\textwidth}
\begin{tabular}{lcccccc}
\hline
Backbone & TDH & UPL & Regions & Units & Cells & Average\\
\hline
Swin-Unetr~\cite{hatamizadeh2021swin} & & & 41.37 & 75.47 & 57.19 & 62.38 \\
Swin-Unetr~\cite{hatamizadeh2021swin} & \checkmark & & 68.55 & 82.70 & 49.90 & 70.97 \\
\hline
Omni-Seg~\cite{deng2023omni} & & & 72.40 & 83.98 & 66.29 & 76.66 \\
Omni-Seg~\cite{deng2023omni} & \checkmark & & 72.43   & 86.39 & 66.49 & 77.89 \\
PrPSeg (Ours) & \checkmark & \checkmark & \textbf{72.51} & \textbf{86.58} & \textbf{68.97} & \textbf{78.66}\\
\hline
\end{tabular}
\end{adjustbox}
\end{center}
\text{*TDH is Token-based Dynamic Head}\\
\text{*UPL is Universal Proposition Learning}\\
\vspace{-7mm}
\label{tab:differentdesigns} 
\end{table}

\section{Conclusion}
\label{sec:conclusion}
In this work, we have developed PrPSeg, a token-based dynamic segmentation network, specifically crafted to facilitate panoramic renal pathology segmentation by effectively modeling the spatial interconnections among diverse anatomical structures. This innovative approach enables the consistent use of the same architectural framework amidst dataset expansions and introduces a universal proposition matrix. This matrix adeptly transforms intricate anatomical relationships into computational modeling paradigms. Furthermore, we have introduced a novel anatomical loss function, integrating these spatial relationships into our model's training regimen through semi-supervised learning. To the best of our knowledge, our algorithm is the first to achieve comprehensive panoramic segmentation in the domain of renal pathology. The integration of token-based dynamic head design alongside our universal proposition learning strategy, which meticulously maps anatomical relationships into the realm of engineering programming, enhances the model’s efficacy in all-encompassing segmentation. 

\section{Acknowledgement}
This research was supported by NIH R01DK135597(Huo), DoD HT9425-23-1-0003(HCY), NIH NIDDK DK56942(ABF). This work was also supported by Vanderbilt Seed Success Grant, Vanderbilt Discovery Grant, and VISE Seed Grant. This research was also supported by NIH grants R01EB033385, R01DK132338, REB017230, R01MH125931, and NSF 2040462. We extend gratitude to NVIDIA for their support by means of the NVIDIA hardware grant. 
{
    \small
    \bibliographystyle{ieeenat_fullname}
    \bibliography{main}
}
\pagebreak
\clearpage
\setcounter{page}{1}





\crefname{section}{Sec.}{Secs.}
\Crefname{section}{Section}{Sections}
\Crefname{table}{Table}{Tables}
\crefname{table}{Tab.}{Tabs.}



\def\paperID{15573} 
\def\confName{CVPR}
\def\confYear{2024}


\thispagestyle{empty}

\appendix
\section{Appendix}

\subsection{Evaluation with Classes Extension}
\label{newclass}
We provide an ablation study for two data extension scenarios: (1) adding 3 new sub-types; (2) introducing 4 new objects to our dataset. The proposed PrPSeg method is flexible to extend new classes by merely updating tokens and the adaptable proposition matrix, without changing the backbone network (\Cref{fig.pipeline}). All models are trained for 30 epochs on the 15-class dataset using the same codebase and experimental settings as those described in the main manuscript. In \Cref{tab:dataextension}, PrPSeg demonstrated superior performance compared to baseline methods across all seven new classes, maintaining the trend observed in \Cref{tab:differentdesigns} of the manuscript.


\let\oldthetable\thetable
\renewcommand{\thetable}{S\arabic{table}} 
\renewcommand{\thefigure}{S\arabic{figure}} 
\renewcommand{\theequation}{S\arabic{equation}} 

\begin{table}[bth]
\caption{
Ablation study with 7 extended classes. Dice similarity coefficient scores (\%) are reported.}
\begin{adjustbox}{width=0.45\textwidth}
\begin{tabular}{lcc|cccccccc}
\toprule
\multirow{1}{0.8in}{Method} & \multirow{1}{0.2in}{TDH}  & \multirow{1}{0.2in}{UPL} & \multicolumn{3}{c}{Regions} & \multicolumn{3}{c}{Functional units}  & \multirow{1}{0.2in}{Cells} & \multirow{1}{0.2in}{Mean} \\
\cmidrule(lr){4-6}
\cmidrule(lr){7-9}
 & & & Inn. Cor.& Mid. Cor. & Out. Cor. & Art. & PTC & MV & Smooth. & \\
\midrule
Swin-Unetr & &  & 34.54 & 31.26 & 34.20 & 53.92 &57.46  &55.03  & 60.18 & 49.66\\
Swin-Unetr  & \checkmark & & 45.25 & 41.89 & 70.22 & 52.27 & 60.95 & 52.17 & 62.45  & 56.67 \\
\midrule
Omni-Seg  & & &39.84 & 43.98 & 70.96 & 47.33 & 63.23  & 48.67 & 56.91 & 53.86 \\
Omni-Seg  & \checkmark &  & 51.16 & 46.46 & 70.53 & 57.89 & 62.63 & 61.93  & 64.41 & 60.43\\
PrPSeg (Ours)  & \checkmark & \checkmark & \textbf{52.69} & \textbf{49.86} & \textbf{71.13} & \textbf{59.51} & \textbf{64.74} & \textbf{63.09} &  \textbf{64.91} & \textbf{61.74} \\
\bottomrule
\end{tabular}
\end{adjustbox}
\text{*TDH is Token-based Dynamic Head}
\text{*UPL is Universal Proposition Learning}
\label{tab:dataextension}
\end{table}

\subsection{Novelty Clarification}
The contributions of this paper are threefold: (1) A comprehensive universal proposition matrix is proposed to provide a simple and adaptable method to model the predominantly overlooked intricate spatial interrelations and class relationships among objects from clinical knowledge. This proposition matrix allows us to flexibly add unseen new classes via minimal changes (only modify tokens and this matrix); (2) The development of a token-based dynamic head in a single network architecture, improving partial label image segmentation. The backbone of the proposed PrPSeg network remains unchanged when new class tokens are introduced for new datasets, enabling the reuse of model weights on incomplete datasets. (3) The formulation of an anatomical loss function that quantifies the inter-object relationships across the kidney. 

\setlength{\abovecaptionskip}{0pt}
\setlength{\belowcaptionskip}{-20pt}
\begin{figure}
\captionsetup{type=figure}
\includegraphics[width=1.0\linewidth]{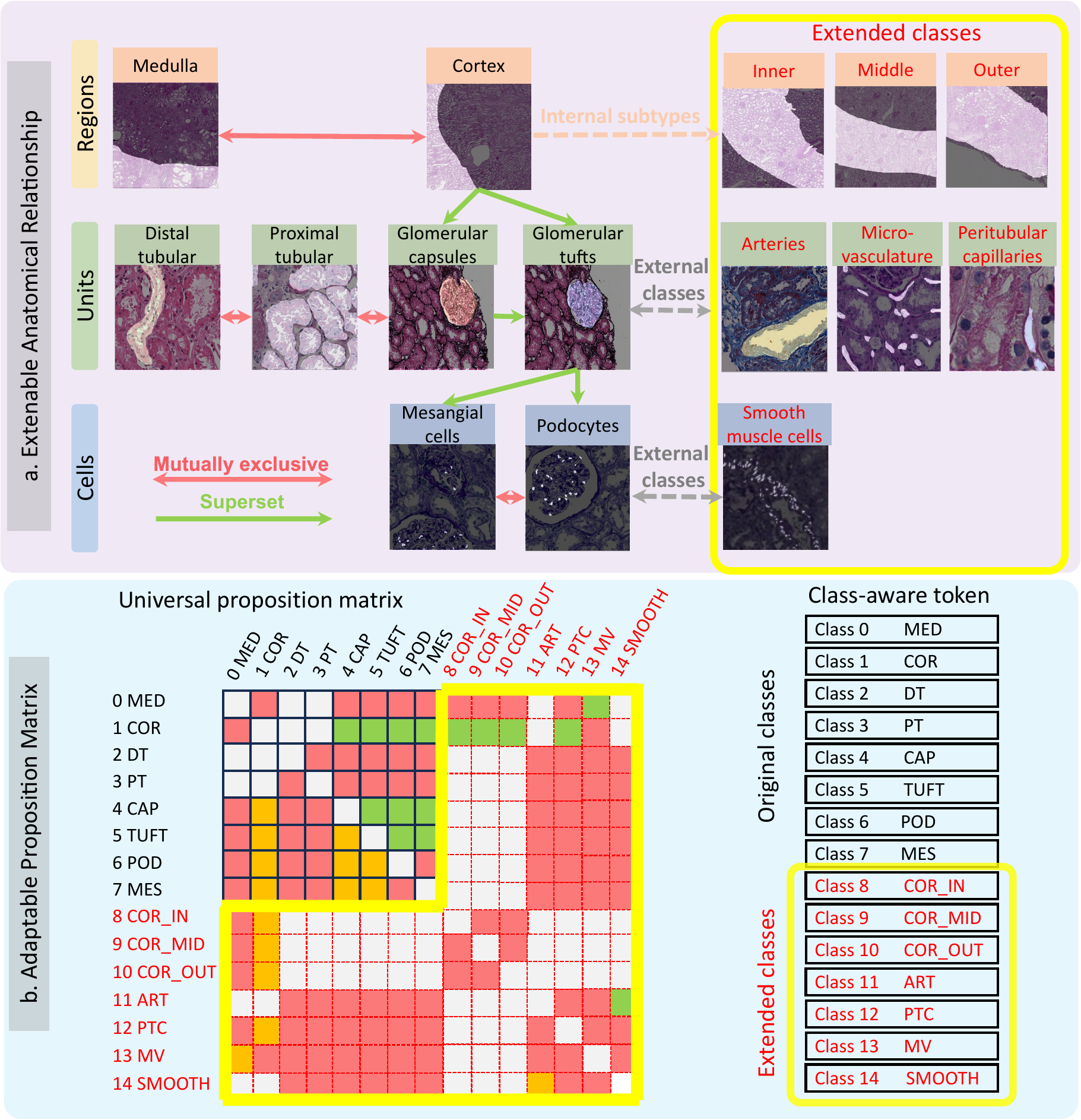}
\captionof{figure}{
\textbf{The innovation of the pipeline when data is extended} -- The proposed PrPSeg method is flexible to extend new classes by merely updating tokens and the adaptable proposition matrix, without changing the backbone network.} 

\label{fig.pipeline} 
\end{figure}

\subsection{Comparison to Related Work and Contributions Beyond Medical Imaging}
Several recent methods that utilize hierarchical information for semantic segmentation~\cite{Li_2022_CVPR,ke2022hsg} or classificationand prediction~\cite{chen2022scaling}. Our method's innovations, beyond previous work, include: (1) Emphasizing pixel-wise anatomical and spatial relationships between objects, rather than solely taxonomy-based relationships (e.g., a glomerulus is located inside the cortex, not merely as a subset or sharing the cortex's morphology); (2) Introducing a hierarchical relationship with class tokens and scale tokens across multiple resolutions (regions at 5$\times$, cells at 20$\times$) to provide greater flexibility between classes and scales, as opposed to a uniform resolution in natural images; (3) Enhancing the extensibility and reusability in model design for data expansion. The proposed method aims to provide a pipeline that is scale-aware, adaptive, and anatomically aware, transitioning from the clinical domain to potential applications in incremental learning and multi-view, multi-scale learning beyond the medical field. 


\end{document}